\documentclass[runningheads]{workshop}
\usepackage[english]{babel}
\usepackage[utf8]{inputenc}
\usepackage{xcolor}
\usepackage{graphicx}  

\newcommand{\beq}{\begin{equation} }
\newcommand{\eeq}{\end{equation}}
\newcommand{\ds}{\displaystyle}
\begin{document}
 
\title*{Compartment model with retarded transition rates}
 
\titretab{Compartment model with retarded transition rates
 }
 
\titrecourt{Compartment model with retarded transition rates}

\author{T\'eo Granger\inst{1}, Thomas Michelitsch$^*$\inst{1}, Bernard Collet\inst{1}, Michael Bestehorn\inst{2} \and Alejandro Riascos\inst{3}}

\index{T Granger }
\index{T Michelitsch}

\auteurcourt{T. Granger, T. Michelitsch \& et al.}
 
\adresse{$^1$ Sorbonne Universit\'e \\
Institut Jean le Rond d'Alembert CNRS UMR 7190 \\
4 place Jussieu, 75252 Paris cedex 05, France  \\
$^2$ Institut für Physik, Brandenburgische 
Technische Universität Cottbus-Senftenberg\\ Erich-Weinert-Straße 1, 03046 Cottbus, Germany 
\\ 
$3$ Departamento de F\'isica, Universidad Nacional de Colombia, Bogot\'a, Colombia}
\email{corresponding $^*$michel@lmm.jussieu.fr}
\maketitle              
\begin{center}
\bf Session:  
Environmental dynamics (hydrology, epidemiology, oceanography, climatology)
\end{center}
\begin{abstract}
Our study is devoted to a four-compartment epidemic model of a constant population of independent random walkers. Each walker is in one of four compartments (S-susceptible, C-infected but not infectious (period of incubation), I-infected and infectious, R-recovered and immune) characterizing the states of health.
The walkers navigate independently on a periodic 2D lattice. Infections occur by collisions of susceptible and infectious walkers.
Once infected, a walker undergoes the delayed cyclic
transition pathway S $\to$ C $\to$ I $\to$ R $\to$ S.
The random delay times between the transitions (sojourn times in the compartments) are drawn from independent probability density functions (PDFs). We analyze the existence of the endemic equilibrium 
and stability of the globally healthy state and derive a condition for the spread of the epidemics which we connect with the basic reproduction number $R_0>1$. We give quantitative numerical evidence that a simple approach based on random walkers offers an appropriate microscopic picture of the dynamics for this class of epidemics.
\end{abstract}

\section{Introduction}
\label{Intro}
The first modern approach of epidemic modelling goes back to the seminal work of
Kermack and Mc Kendrick \cite{KermackMcKendrick} who introduced the first `SIR - compartment type model' (S-I-R standing for the states
susceptible, infected and recovered (immune), respectively. In the meantime, epidemic modelling has become a huge field \cite{Anderson1992,Martcheva2015,Satoras-Vespignani-etal2015,ZhuShenWang2023} (and many others).

Our study is devoted to an epidemic model for a constant population by taking into account four
compartments of individuals characterizing their states of health. Each individual is
in one of the compartments susceptible (S); incubated -- infected yet not infectious (C),
infected and infectious (I), and recovered -- immune (R). An infection is visible only
when an individual is in state I.
Upon infection, an individual performs the transition
pathway S $\to $ C $\to$ I $\to$ R $\to$ S remaining in each compartments C, I, and R for certain
random waiting time $t_C$, $t_I$ , $t_R$, respectively. The waiting times (sojourn times) in each compartment are
independent and drawn from specific probability density functions (PDFs) introducing
memory effects into the model \cite{Granger-et-al-PRE2023,SCIRS-model} generalizing our previous model \cite{Markovian}. 

Based on these assumptions, we introduce first the macroscopic SCIRS
model and derive memory equations for the epidemic evolution involving convolutions (time derivatives of
general fractional type in the Kochubei sense \cite{Kochubei2011}). The classical (memoryless) version of the model is recovered
for exponentially distributed compartment waiting times. For long waiting times drawn from fat-tailed (power-law) distributions, the SCIRS evolution equations
take the form of time-fractional ODEs \cite{Granger-et-al-PRE2023}. 

We obtain formulae for the endemic equilibrium
and a condition of its existence for cases where the waiting time PDFs have existing means.
We analyze the stability of healthy and endemic equilibria and derive conditions of its existence. 

\begin{figure}
\includegraphics[width=0.53\textwidth]{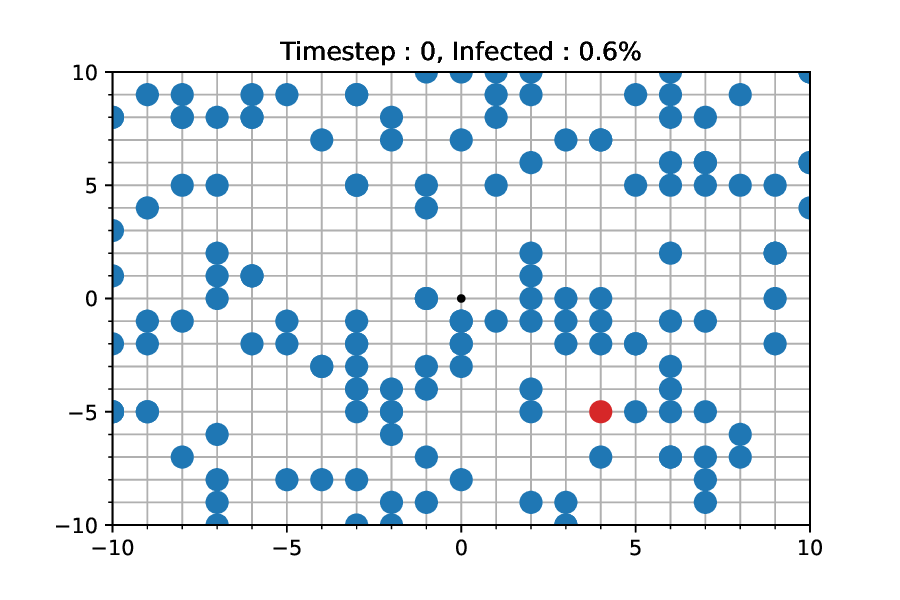} 
\includegraphics[width=0.53\textwidth]{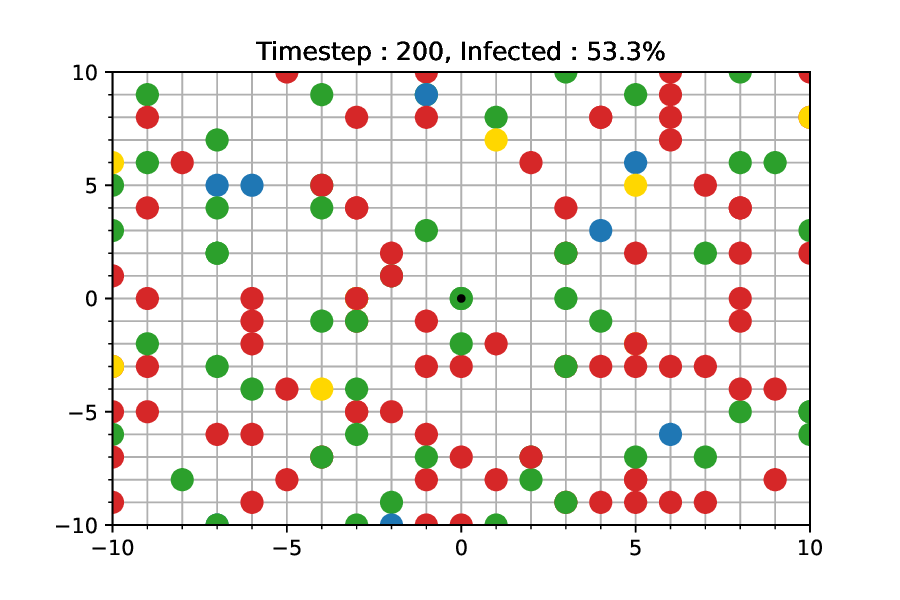}
\caption{Multiple random walkers model: Colors indicate the health states (compartments) of the walkers: S (blue), C (yellow), I (red), R (green). Left plot: Typical initial condition with one infected walker. Right plot: State of epidemic spreading for $t>0$ with Gamma-distributed $t_{C,I,R}$ with parameters given subsequently.}
\label{lattice}
\end{figure}

We implemented a multiple random walker's model into a PYTHON code (which is freely available \cite{SCIRS-model}) where 
$Z$ independent walkers navigate independently on a $N \times N$ periodic (ergodic) square lattice. The initial positions of the walkers on the lattice are random.
In each time increment, the walkers perform simultaneously independent jumps to one of their four neighbor lattice sites with equal probability $\frac{1}{4}$ (simple unbiased walk). Each walker is in one of the compartments S,C,I,R (Fig. \ref{lattice}).
Infections occur with a certain probability only when infectious I walkers meet susceptible S walkers on the same lattice sites. 
Once a walker is infected in this way, he undergoes the above explained cyclic SCIRS transition pathway with random sojourn times $t_{C,I,R}$ in compartments C I R.
We compare the endemic states predicted analytically by the macroscopic model with the numerical
results of the random walk simulations (long time asymptotics of the compartmental populations) and find accordance with high accuracy.

\section{SCIRS model}
\label{Model}
Let $s(t)=\frac{Z_S(t)}{Z}, c(t)=\frac{Z_Ct)}{Z}, j(t)=\frac{Z_I(t)}{Z}, r(t)=\frac{Z_R(t)}{Z}$ be the fractions of the population in the compartments S C I R, corresponding to $Z_{S,C,I,R}(t)$ random walkers in these compartments. We neglect birth and death processes and consider the total number of walkers to be constant $Z=Z_S(t)+Z_C(t)+Z_I(t)+Z_R(t) \gg 1$. We denote with $t_C,t_I,t_R >0 $ the random sojourn times (waiting times) a walker spends in compartments C, I, R, respectively and with ${\cal A}(t)$ the infection rate at time $t$. The infection rate actually contains microscopic (random walk) information on the collisions of I and S walkers and transmission probability of the disease. We introduce a kind of predator-prey model where the I walkers are predators and S walkers the prey, with a simple nonlinear law ${\cal A}(t) = 
{\cal A}(j(t),s(t)) = \beta j(t)s(t)$ where $\beta >0$ denotes a time independent positive constant depending on the probability of infection in a collision of I and S walkers among other random walk characteristics. 
We propose the following evolution equations
\beq
\label{SRIRS-delta}
\begin{array}{clr}
 \ds \frac{d}{dt}s(t) & = \ds  -{\cal A}(t) +  \langle {\cal A}(t-t_C-t_I-t_R) \rangle & \\ \\
\ds \frac{d}{dt}c(t) & =  \ds {\cal A}(t) - \langle {\cal A}(t-t_C) \rangle & \\ \\
\ds \frac{d}{dt}j(t) & = \ds  \langle {\cal A}(t-t_C) \rangle -\langle {\cal A}(t-t_C-t_I)\rangle  & \\ \\
\ds \frac{d}{dt}r(t) & = \ds  \langle {\cal A}(t-t_C-t_I) \rangle - \langle{\cal A}(t-t_C-t_I-t_R) \rangle &
\end{array}
\eeq
where we assume as initial condition the globally healthy state $s(0)=1^{-}$, $j(0)=0^+$ $c(0)=r(0)=0$ where (almost) all walkers are in compartment S. $\langle \ldots\rangle$ indicates averaging over the contained random variables $t_C,t_I,t_R$. Since the total population is constant,
one of the four equations is redundant, however we write them here all for clarity.
To perform this average, we assume the compartment sojourn times to be mutually independent and drawn from causal\footnote{i.e. $K_{C,I,R}(\tau)=0$ for $\tau <0$ reflecting strict positivity of $t_{C,I,R}$.} probability density functions (PDFs) $K_{C,I,R}(\tau)$ such that
$$
Prob(t_{C,I,R}\in [\tau,\tau+{\rm d}\tau]) =  K_{C,I,R}(\tau){\rm d}\tau , \hspace{1cm} t_{C,I,R} >0
$$
indicating the probabilities that $t_{C,I,R} \in [\tau,\tau+{\rm d}\tau]$.
Then the following averaging rule applies
\beq
\label{averag_rule}
\langle f(t_{C,I,R}) \rangle = \int_0^{\infty}f(\tau) K_{C,I,R}(\tau){\rm d}\tau 
\eeq
for suitable functions $f(\tau)$ to perform in (\ref{SRIRS-delta}) the average over the independent random variables $t_{C,I,R}$. This operation together with causality of the involved functions takes us to the explicit convolutional representation of the SCIRS evolution equations \cite{Granger-et-al-PRE2023}
\beq
 \label{SCIRS-model}
 \begin{array}{clr}
\ds  \frac{d}{dt}s(t) & = \ds  - {\cal A}(t)   +  ({\cal A} \star K_C \star K_I \star K_R)(t) & \\  \\
\ds \frac{d}{dt}c(t) & = \ds {\cal A}(t) - ({\cal A} \star K_C)(t) \\ \\
\ds  \frac{d}{dt}j(t) & =  \ds ({\cal A} \star K_C)(t) - ({\cal A} \star K_C \star K_I)(t) \\ \\
 \ds  \frac{d}{dt}r(t) & = \ds ({\cal A} \star K_C \star K_I)(t) -  
  ({\cal A} \star K_C \star K_I\star  K_R)(t) 
 \end{array}
\eeq
where $(a \star b)(t) = \int_0^t a(\tau)b(t-\tau){\rm d}\tau$ stands for convolution of the causal functions $a(t),b(t)$.
The interpretation of Eqs. (\ref{SRIRS-delta}), (\ref{SCIRS-model}) is as follows. The transition rate ${\cal A}(t)$ out of compartments S into C is the rate of new infections at time $t$ (see first and second lines in (\ref{SRIRS-delta})). Then the randomly delayed transitions out of C into I (individuals who fall sick) have the rate $\langle {\cal A}(t-t_C)\rangle $ coming from infections at $t-t_C$. Further, the term $\langle {\cal A}(t-t_C-t_I)\rangle$ captures transitions out of I into R (healed individuals). Finally, $\langle {\cal A}(t-t_C-t_I-t_R)\rangle$ is the rate of transitions out of R into S (individuals loosing their immunity at time $t$) closing the cyclic infection pathway. Be aware that ${\cal A}(\tau)$ is causal, i.e. vanishing for negative arguments $\tau$.

The randomly delayed transitions generally introduce memory effects, where exponentially distributed waiting times
correspond to kernels $K(\tau)=  \xi e^{-\xi \tau}$ representing the memoryless (Markovian) case.
In our study, we mainly focus on Gamma-distributed sojourn times with PDFs
$K_{\alpha,\xi}(\tau) = \xi^{\alpha}\frac{\tau^{\alpha-1}}{\Gamma(\alpha)} e^{-\xi \tau}$ ($\alpha, \xi>0$ indicating shape and rate parameters, respectively) with sufficient flexibility to capture a wide range of behaviors, including the memoryless case with exponential PDF for $\alpha=1$ and the limit of sharp waiting times $\delta$-distributed $t_{C,I,R}$ for $\xi \to \infty$ while the mean waiting time $\frac{\alpha}{\xi}$ is kept constant.
Laplace transforming the SCIRS  equations leads to the following endemic states (large time asymptotic compartmental fractions) \cite{Granger-et-al-PRE2023}
\beq
\label{endem}
\begin{array}{clr} 
\ds S_e &= \ds  \frac{1}{R_0} & \\ \\
\ds  C_e & = \ds \frac{R_0-1}{R_0} \frac{\langle t_C \rangle }{\langle T \rangle}  & \\ \\
\ds J_e & = \ds \frac{R_0-1}{R_0} \frac{\langle t_I \rangle}{\langle T \rangle}  &\\ \\
\ds R_e & = \ds  \frac{R_0-1}{R_0} \frac{\langle t_R \rangle}{\langle T \rangle} & 
\end{array} \hspace{1cm} \ds  \langle T \rangle = \langle t_C+t_I+t_R \rangle , \hspace{1cm} R_0=\beta \langle t_I \rangle
\eeq
for existing mean sojourn times $ \langle t_{C,I,R} \rangle = \int_0^{\infty} \tau K_{C,I,R}(\tau){\rm d}\tau < \infty$.
The endemic equilibrium
exists solely for $R_0 = \beta \langle t_I \rangle >1$ where we interpret $R_0$ as basic reproduction number (average number of infected walkers produced by one single initially infected walker during his average time of infection $\langle t_I \rangle$). The endemic equilibrium only depends on $R_0$ and the means $\langle t_{C,I,R}\rangle $.

This interpretation of $\beta \langle t_I \rangle$ as basic reproduction number becomes more clear when we consider in the second equation of (\ref{SCIRS-model}) the number of new infections per time unit at $t=0$ caused by a single I walker $Z_I(0)=1$, namely
\beq
\label{R0_def_relation}
\frac{dZ_c(t)}{dt}\big|_{t=0} = Z \beta s(t)j(t)\big|_{t=0} = \frac{\beta}{Z}Z_S(t)Z_I(t)\big|_{t=0} = \beta \frac{Z-1}{Z} \to \beta ,\hspace{1cm} Z \gg 1
\eeq
where $\langle {\cal A}(t-t_C) \rangle\big|_{t=0} = 0$ because of causality.
Taking into account that the first infected walker can infect other walkers during the average time of infection $\langle t_I \rangle$, he can indeed infect (in a first order approximation) in the average $\langle t_I \rangle \frac{dZ_c(t)}{dt}\big|_{t=0} = \beta \langle t_I \rangle =R_0$ susceptible walkers.
\vspace{1cm}
\section{Discussion and results}
\label{discussion}

We consider in the following the role of jump length (short- and long-range navigation) in the lattice on the epidemic spreading.

\begin{figure}
\centerline{\includegraphics[width=0.53\textwidth]{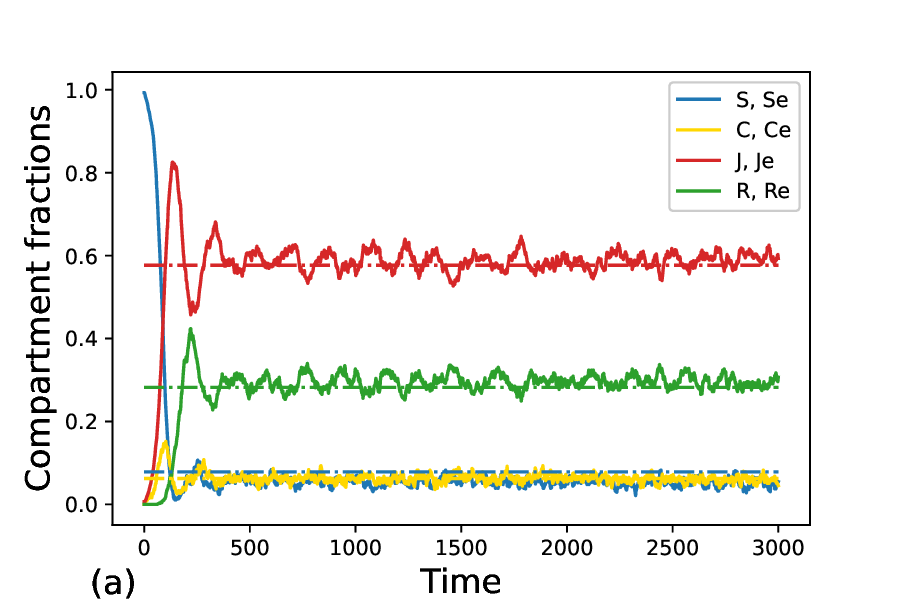}
\includegraphics[width=0.53\textwidth]{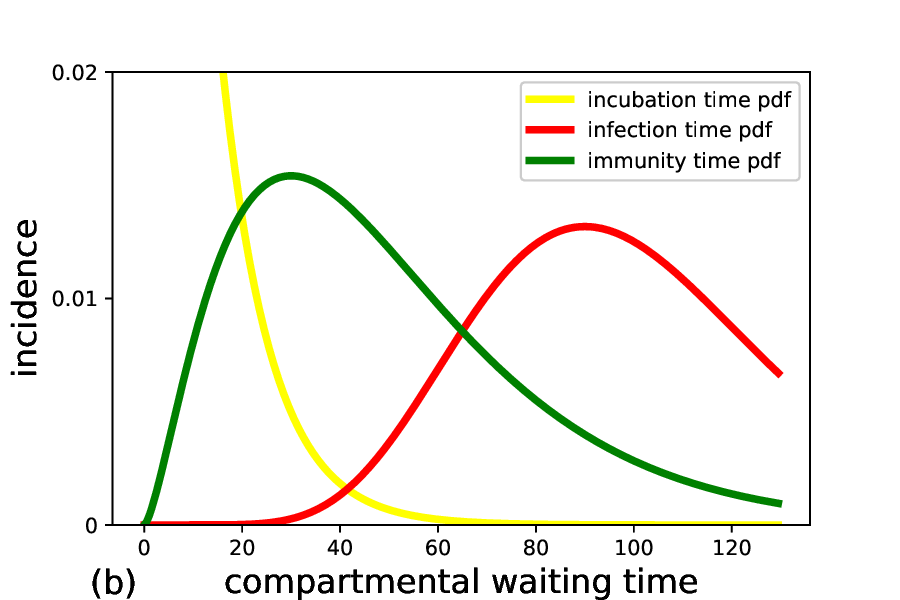}}
\caption{(a) Random walk simulation where all walkers perform short-range steps to neighbor nodes for Gamma-distributed compartment sojourn times with $ \langle t_{C} \rangle = 10,  \langle t_{I} \rangle= 100,  \langle t_{R} \rangle = 50$ and $\xi_C=0.1, \xi_I=0.1, \xi_R=0.05$. Dashed lines indicate endemic equilibrium values. (b) Gamma waiting-time PDFs of incubation $t_C$ (yellow), infection $t_I$ (red), and immunity $t_R$ (green) times. The remaining parameters are $Z=150$ walkers, with $N\times N$ lattice ($N=21$), and infection probability in a collision is $0.8$. The initial condition is one infected and $Z-1$ susceptible walkers. }
\label{evol1}
\end{figure}

In Fig. \ref{evol1}(a) is drawn the SCIRS evolution where all walkers perform short-range steps to neighbor lattice sites. In each time step, we count the compartmental population where we average 5 equivalent random walk runs with the same parameters but different random numbers (PYTHON seeds). The compartmental waiting times $t_{C,I,R}$ in the random walk simulations are determined as random numbers drawn from Gamma-distributions (specified in Fig \ref{evol1}(b)).
The dashed lines indicate the numerically
determined endemic values (by counting the compartmental populations $Z_{S,C,I,R}(t)$) and are obtained as $S_e\approx 0.08$, $C_e\approx 0.06$, $J_e \approx 0.58$
$R_e\approx 0.27$ (with $S_e+C_e+J_e+R_e=1$) and basic reproduction number $R_0 = 1/S_e \approx 12.75$. Inspection of these numerical values shows that the ratios $(C_e : J_e : R_e) = (\langle t_C \rangle :  \langle t_I \rangle :  \langle t_R \rangle) \approx (1 : 10 : 5)$ are in excellent agreement with the ratios predicted from the analytically derived Eqs. (\ref{endem}) for the endemic equilibrium.
Increasing the observation time improves the agreement. 
This shows that a simple random walk approach offers an appropriate microscopic picture of the macroscopic SCIRS dynamics (evolution Eqs. (\ref{SRIRS-delta}), (\ref{SCIRS-model})).

We refer to \cite{Granger-et-al-PRE2023} for extensive discussions and case studies which further confirm the validity of (\ref{endem}) where animated simulations (videos) can be consulted in the supplementary materials \cite{SCIRS-model}.

\begin{figure}
\centerline{\includegraphics[width=0.53\textwidth]{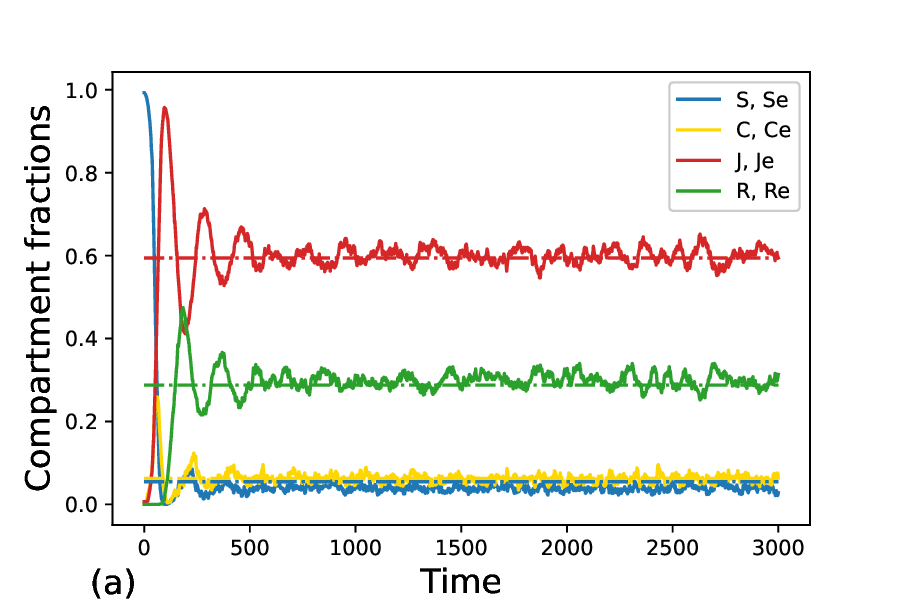}
\includegraphics[width=0.53\textwidth]{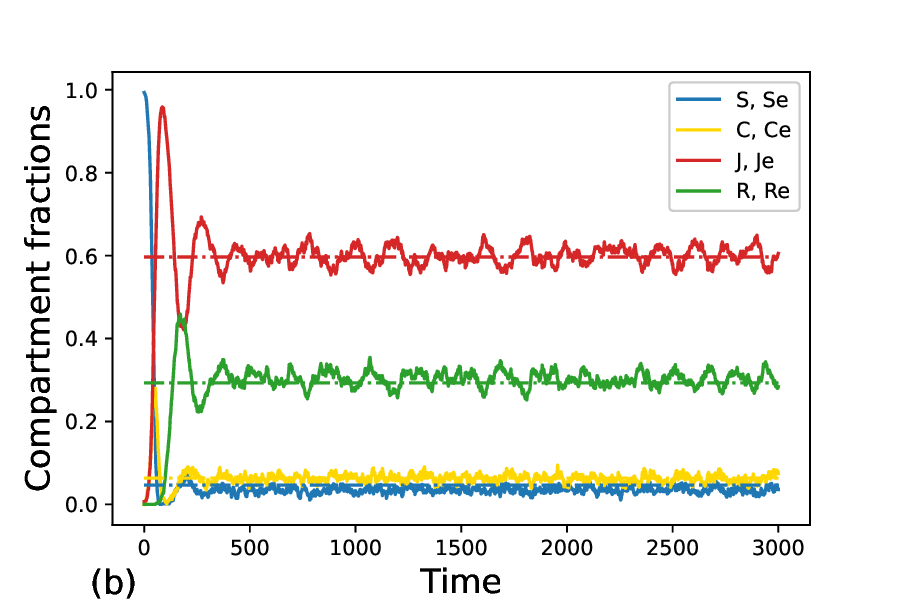}}
\caption{Time evolution for (a) 40\% (b)  90\% superspreaders. All other parameters and waiting time distributions are the same as in Fig. \ref{evol1}.}
\label{evolution2}
\end{figure}

In the simulation runs of Fig. \ref{evolution2} we choose all parameters identically as in Fig. \ref{evol1}, however we allow a certain fraction of walkers which we refer to as ``superspreaders'' to perform at any time-increment long-range jumps to any lattice site of the lattice with equal probability. The remaining walkers jump with short steps to neighbor sites. One can see that the basic reproduction number monotonously increases as the fraction of superspreaders increase (from Figs. \ref{evolution2}(a) to (b)). The walks of superspreaders correspond to navigation on a fully connected (graph) architecture. In Fig. \ref{evolution2}(a) with $40\%$ superspreaders the basic reproduction number is $R_0\approx 18.25$ and is considerably increased compared to the value $12.75$ of Fig. \ref{evol1}(a) ($0\%$ superspreaders) thus the endemic value $J_e \approx 0.6$ is in Figs. \ref{evolution2}(a,b) slightly higher than without superspreaders. When we increase the fraction of superspreaders to $90\%$ (Fig. \ref{evolution2}(b)), the basic reproduction number further increases to $R_0 \approx 21.48$. One can see in Figs. \ref{evolution2}(a,b) that the endemic values only slightly change, however the first infection wave becomes 
much more pronounced, reaching very high maximum values $j_{max} \approx 0.97$. On the other hand, the evolutions with $40\%$ and $90\%$ differ only by their oscillatory behaviors. Higher fractions of superspreaders seem to have the effect that fluctuations around the endemic values have smaller amplitudes.
A conclusion from these observations is the recommendation to decision makers to avoid long-range navigation of individuals in epidemic contexts in order to mitigate the first infection wave. On the other hand,
such a measure seems to have only very little impact on the endemic equilibrium.

\section{Conclusions}
\label{Conclusions}
In the present letter, we investigated epidemic spreading on a two-dimensional periodic lattice with a cyclic SCIRS infection pathway where the transitions occur with random delay. We focused here on Gamma-distributed delay times.
In a follow-up project, it would be desirable to have a microscopic theory connecting the phenomenological infection rate ${\cal A}(t)$ with random walk characteristics such as collision rate, transmission probability when walkers meet, and the topology of the network if it is more complex than a  simple two-dimensional lattice.
In this context an interesting direction is the epidemic spreading in complex small or large world networks where the complexity of the network architecture may have a crucial impact on the epidemic spreading \cite{Tobepublished}.
For future research, it would be interesting to explore whether Eqs. (\ref{SRIRS-delta}) or similar systems with simple non-linear infection rates may exhibit chaotic attractors \cite{OERoessler1076} as endemic states for certain sets of parameters and waiting time distributions.

\vspace{1cm}

\bibliographystyle{plain}

\end{document}